\documentclass[prx,twocolumn,floatfix,a4paper,superscriptaddress]{revtex4}
\usepackage{bm,color,graphicx,amsmath,txfonts}

\usepackage{hyperref}

\hypersetup{
 colorlinks=true,   
 linkcolor=blue,  
 citecolor=blue, 
 urlcolor =blue    
}

\newcommand{\CC}{{\cal C}}

\newcommand{\GG}{{\cal G}}

\newcommand{\FF}{{\cal F}}

\begin{document}

\title{Quantum network with magnonic and mechanical nodes}

\author{Jie Li}\thanks{jieli007@zju.edu.cn}
\affiliation{Interdisciplinary Center of Quantum Information, State Key Laboratory of Modern Optical Instrumentation, and Zhejiang Province Key Laboratory of Quantum Technology and Device, Department of Physics, Zhejiang University, Hangzhou 310027, China}
\author{Yi-Pu Wang}\thanks{yipuwang@zju.edu.cn}
\affiliation{Interdisciplinary Center of Quantum Information, State Key Laboratory of Modern Optical Instrumentation, and Zhejiang Province Key Laboratory of Quantum Technology and Device, Department of Physics, Zhejiang University, Hangzhou 310027, China}
\author{Wei-Jiang Wu}
\affiliation{Interdisciplinary Center of Quantum Information, State Key Laboratory of Modern Optical Instrumentation, and Zhejiang Province Key Laboratory of Quantum Technology and Device, Department of Physics, Zhejiang University, Hangzhou 310027, China}
\author{Shi-Yao Zhu}
\affiliation{Interdisciplinary Center of Quantum Information, State Key Laboratory of Modern Optical Instrumentation, and Zhejiang Province Key Laboratory of Quantum Technology and Device, Department of Physics, Zhejiang University, Hangzhou 310027, China}
\author{J. Q. You}
\affiliation{Interdisciplinary Center of Quantum Information, State Key Laboratory of Modern Optical Instrumentation, and Zhejiang Province Key Laboratory of Quantum Technology and Device, Department of Physics, Zhejiang University, Hangzhou 310027, China}

\begin{abstract}
A quantum network consisting of magnonic and mechanical nodes connected by light is proposed. Recent years have witnessed a significant development in cavity magnonics based on collective spin excitations in ferrimagnetic crystals, such as yttrium iron garnet (YIG). Magnonic systems are considered to be a promising building block for a future quantum network. However, a major limitation of the system is that the coherence time of the magnon excitations is limited by their intrinsic loss (typically in the order of 1 $\mu$s for YIG). Here, we show that by coupling the magnonic system to a mechanical system using optical pulses, an arbitrary magnonic state (either classical or quantum) can be transferred to and stored in a distant long-lived mechanical resonator. The fidelity depends on the pulse parameters and the transmission loss. We further show that the magnonic and mechanical nodes can be prepared in a macroscopic entangled state. These demonstrate the quantum state transfer and entanglement distribution in such a novel quantum network of magnonic and mechanical nodes. Our work shows the possibility to connect two separate fields of optomagnonics and optomechanics, and to build a long-distance quantum network based on magnonic and mechanical systems. 
\end{abstract}

\date{\today}
\maketitle

\section{Introduction}

Hybrid quantum systems, composed of distinct physical systems with complementary functionalities, provide diverse novel platforms and promising opportunities for applications in quantum technologies, quantum-information processing and quantum sensing~\cite{hybrid0,hybrid1,hybrid2}. It merits our particular attention that, during the past decade a rapid and significant progress has been made in the field of cavity magnonics, based on coherently coupled microwave cavity photons and collective spin excitations in the ferrimagnetic material of yttrium iron garnet (YIG)~\cite{NakaRev,NatRev,RMP,S1,S2,S3,S5,S6,S7,S8,S9,S9b,S10,S11,S12,S13,S14,S15,S16,S17,S18,S19,S20,S21,S22,S23,S24,S25,S26,S27,S28,S29}. Cavity magnonics has now become a new platform for the study of strong interactions between light and matter, in the context of cavity QED with magnons.  As one of the main advantages, the magnonic system shows an excellent ability to coherently interact with diverse quantum systems, including microwave~\cite{S1,S2,S3} or optical photons~\cite{S7,S8,S9}, phonons~\cite{S9b,S17,S26,S29}, and superconducting qubits~\cite{S5,S10,S23,S27}. Hybrid cavity magnonic systems promise potential applications in quantum-information processing~\cite{NakaRev}, quantum sensing~\cite{Yan,Ruoso,Ali}, and in searching dark-matter axions~\cite{DM}, to name a few.

In this paper, we show the potential to build a quantum network~\cite{Kimble,Wehner} based on magnonic systems in view of their aforementioned excellent properties. A future quantum network could be constructed based on single atoms in optical cavities~\cite{Rempe}, or atomic ensembles following the Duan-Lukin-Cirac-Zoller protocol~\cite{DLCZ}, or trapped atomic ions~\cite{Duan}, etc. Compared to these platforms as quantum nodes, where the atomic energy levels are fixed, a major advantage of magnonic systems lies in the fact that their resonance frequencies can be continuously adjusted by altering the external magnetic field. This offers a large flexibility to couple to different quantum systems, like superconducting qubits, photons, and phonons~\cite{NakaRev}. Therefore, a quantum network based on magnonic systems shows its unique advantages.  However, a major obstacle for such a magnon-based quantum network is that its coherence time is limited by its intrinsic loss (typically with damping rate $\gamma_m/2\pi \,\,{\sim} \,1$ MHz), and is in the order of 1 $\mu$s for YIG. The coherence time can indeed be significantly extended by transferring the magnonic quantum state to the mechanical mode (i.e., the vibrational phonon mode) of the {\it same} YIG ferrimagnet~\cite{Jie19,JieQST,Jing,Sarma}, which can act as a long-lived quantum memory~\cite{Simon,Simon2}. However, this local operation via magnomechanics does not allow to build a quantum network with its nodes distributed in a long distance. 

Here, we show that this obstacle can be eliminated by using light, an optimal candidate for transmitting quantum information over a long distance, through which the magnonic system is connected to a distant mechanical system. We, for the first time, prove that light can connect two separate fields of optomagnonics and optomechanics, and be used to accomplish some basic functions of a quantum network, such as quantum state transfer and entanglement distribution among different nodes of the network~\cite{Kimble,Rempe,cirac}. The quantum network with magnonic and mechanical nodes combines the advantages of both systems, i.e., the great magnonic compatibility and tunability as well as the long mechanical coherence time. Remarkably, a recent optomechanical experiment has demonstrated a mechanical coherence time longer than 100 ms~\cite{Albert}. 
Specifically, we show that an arbitrary magnonic state, either quantum or classical, can be transferred to a distant mechanical resonator by using optical pulses to successively activate the optomagnonic and optomechanical anti-Stokes processes. This allows the transfer of the magnonic state to the anti-Stokes optical pulse, and then the mapping of the pulse state to the mechanical mode. The magnonic state can be stored in the mechanical mode within its coherence time and retrieved by sending a weak red-detuned read pulse to the optomechanical cavity, of which the output field carries the magnonic state. We study the fidelity in this magnon-to-phonon state transfer process, and show its dependence on the system parameters, e.g., the pulse strengths and durations, and the transmission loss.

We further show that the magnonic and mechanical nodes can be prepared in a macroscopic entangled state by using pulses to successively activate the optomagnonic Stokes and optomechanical anti-Stokes processes. The former process realizes a two-mode squeezed vacuum state of the magnons and the pulse, and the latter process maps the pulse state onto the mechanical mode, thus establishing a nonlocal entangled state between the magnonic and mechanical nodes.  

The paper is organized as follows. In Sec.~\ref{bas}, we introduce some basic interactions in cavity optomagnonics and optomechanics, which are necessities for realizing our protocol. In Sec.~\ref{transf}, we show how to operate the system via optical pulses such that an arbitrary magnonic state can be transferred to a distant long-lived mechanical resonator. We further analyse the fidelity in this state transfer process and show its dependence on the system parameters, especially on those related to optical pulses. We then study the effects of the transmission loss of the pulse on the transferred mechanical state and the fidelity. In Sec.~\ref{entang}, we show how to prepare a nonlocal macroscopic entangled state between the magnonic and mechanical nodes, and provide a strategy to detect it. Finally, we draw the conclusions in Sec.~\ref{conc}.

\section{Basic interactions in optomagnonics and optomechanics}
\label{bas}

We start with the introduction of the basic interactions in cavity optomagnonics and optomechanics that are key building blocks for realizing our protocol. These include the optomagnonic (optomechanical) two-mode squeezing and beamsplitter interactions that are used for realizing the entangling and state-swap operations, respectively. We explicitly show how to operate the two subsystems to realize these interactions and provide their effective Hamiltonians. With the successful implementation of these local operations, we prove in the next section that a remote quantum network based on magnonic and mechanical systems can be built using optical pulses.

\subsection{Magnon-induced Brillouin light scattering in optomagnonics}

\begin{figure}[t]
\hskip0.0cm\includegraphics[width=0.85\linewidth]{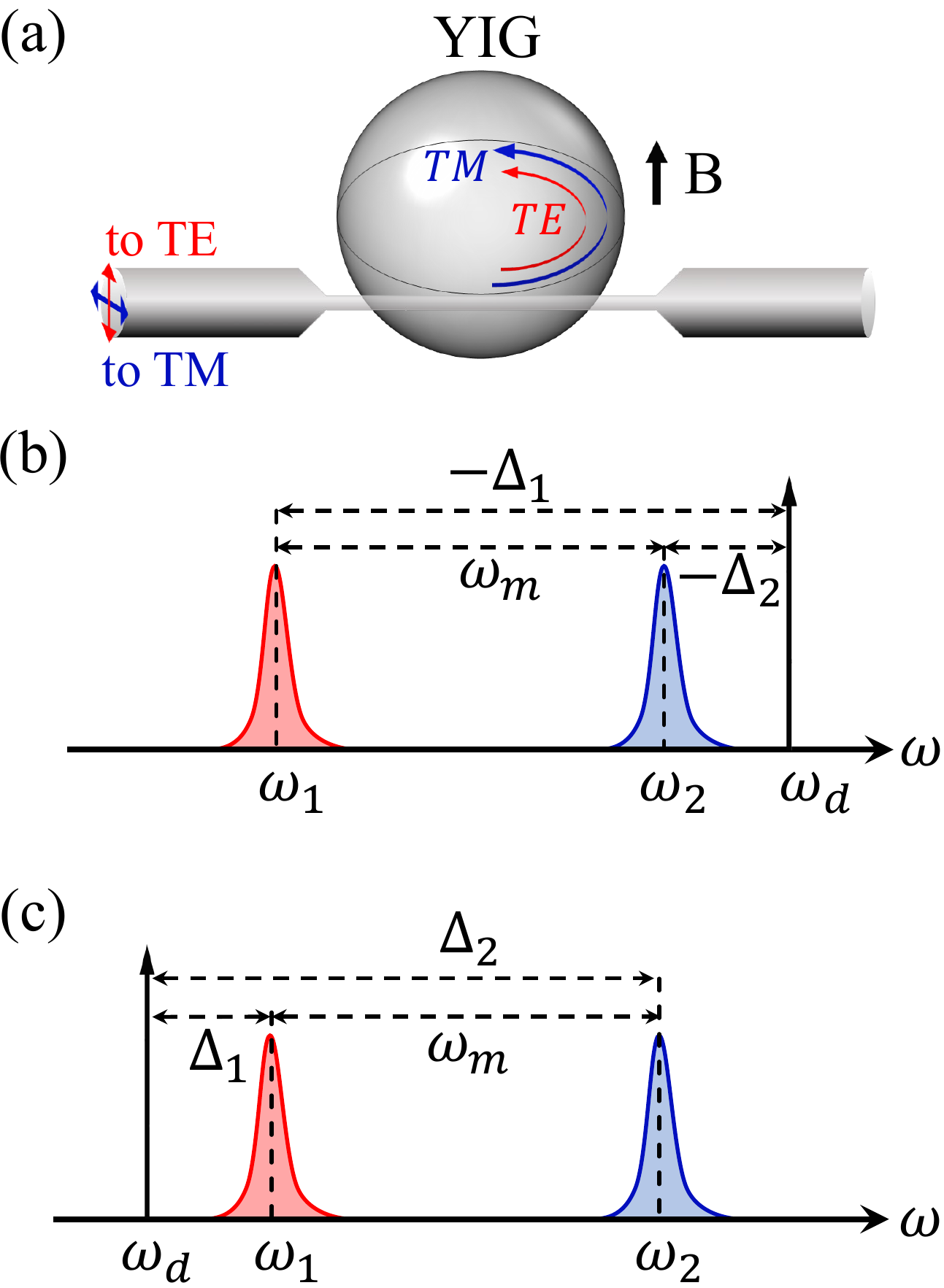} 
\caption{(a) An optomagnonic system of a YIG sphere supporting two WGMs and a magnon mode. (b) Mode frequencies of the optomagnonic Stokes BLS. (c) Mode frequencies of the optomagnonic anti-Stokes BLS. }
\label{fig1}
\end{figure}

We consider a cavity optomagnonic system of a YIG sphere~\cite{S7,S8,S9} that simultaneously supports a magnetostatic mode of magnons and whispering gallery modes (WGMs) of optical photons, as depicted in Fig.~\ref{fig1}(a). The photons in a WGM are scattered by the lower-frequency magnons, typically in GHz~\cite{S7,S8,S9}, yielding sideband photons with their frequency shifted by the magnon frequency. This process is known as the magnon-induced Brillouin light scattering (BLS). When the scattered photons go into another WGM (the so-called triple-resonance condition), the BLS scattering probability is then maximized. This triple resonance can be conveniently achieved by tuning the magnon frequency realized by altering the strength of the bias magnetic field. Owing to the selection rule~\cite{GEB17,Papa,UsamiNJP,Haigh18} imposed by the conservation of the angular momenta of WGM photons and magnons, the BLS shows a pronounced asymmetry in the Stokes and anti-Stokes scattering strengths. This asymmetry is the basis for realizing the proposals for preparing macroscopic quantum states of magnons in optomagnonics~\cite{GEB18,SVK19,Zhou,Xie,Jie21,Heqy}.

The magnon-induced BLS is intrinsically a three-wave process, which can be described by the Hamiltonian
\begin{equation}
H= H_0 + H_{\rm int} + H_{d},
\end{equation}
where $H_0$ is the free Hamiltonian of two WGMs and a magnon mode
\begin{equation}
H_0/\hbar= \omega_1 a_1^{\dag} a_1 + \omega_2 a_2^{\dag} a_2 + \omega_m m^{\dag} m,
\end{equation}
with $a_j$ and $m$ ($a_j^{\dag}$ and $m^{\dag}$, $j=1,2$) being the annihilation (creation) operators of the WGMs and magnon mode, respectively, and $\omega_i$ ($i=1,2,m$) being their resonance frequencies, which satisfy the relation $\omega_m \ll \omega_j$ and $|\omega_1-\omega_2|=\omega_m$, imposed by the conservation of energy in the BLS. The interaction Hamiltonian $H_{\rm int}$ of the three modes is given by
\begin{equation}
H_{\rm int}/\hbar= G_0 \big(a_1^{\dag} a_2 m^{\dag} + a_1 a_2^{\dag} m \big),
\end{equation}
where $G_0$ is the single-photon coupling rate. This coupling is weak due to the large frequency difference between the optical and magnon modes, but it can be significantly enhanced by intensely driving one of the WGMs. The driving Hamiltonian is 
\begin{equation}
H_d/\hbar= i E_j \big(a_j^{\dag} e^{-i\omega_d t} - a_j e^{i\omega_d t} \big),
\end{equation}
where $E_j=\!\sqrt{P_j \kappa_j^{e}/\hbar \omega_d}$ is the coupling strength between the $j$th WGM (with external decay rate $\kappa_j^{e}$) and the driving field (with frequency $\omega_d$ and power $P_j$). To maximize the BLS scattering probability, we resonantly pump either the WGM $a_1$ or $a_2$~\cite{S7,S8,S9} (i.e., $\omega_d = \omega_1$ or $\omega_2$) to {\it selectively} activate the anti-Stokes or Stokes scattering, which is responsible for the optomagnonic state-swap or two-mode squeezing interaction. Note that the selection rule also causes different optical polarizations of the two WGMs. Without loss of generality, we assume $a_2$ ($a_1$) mode to be the TM (TE) mode of a certain WGM orbit, and $\omega_{2({\rm TM})}>\omega_{1({\rm TE})}$ due to the geometrical birefringence of the WGM resonator~\cite{S7,Jie21}. It is worth noting that this is true for WGMs with the same angular momentum, but when the optical angular momentum changes the dominant optomagnonic coupling occurs between WGMs satisfying $\omega_{{\rm TM}}<\omega_{{\rm TE}}$~\cite{GEB17,UsamiNJP,Haigh18}.

The Hamiltonian $H$ takes a compact form in the frame rotating at the drive frequency $\omega_d$, which is
\begin{equation}
\begin{split}
H/\hbar &= \Delta_1 a_1^{\dag} a_1 + \Delta_2 a_2^{\dag} a_2 + \omega_m m^{\dag} m  \\
&+ G_0 \big(a_1^{\dag} a_2 m^{\dag} + a_1 a_2^{\dag} m \big) + i E_j \big( a_j^{\dag} - a_j \big),
\end{split}
\end{equation}
where $\Delta_j = \omega_j - \omega_d$ ($j=1,2$) is the cavity-drive detuning. We now consider the case where mode $a_2$ is resonantly pumped by a strong optical field, i.e., $\Delta_2=0$, and thus $\Delta_1=-\omega_m$, cf. Fig.~\ref{fig1}(b). This can be realized by, e.g., coupling the laser field with a certain polarization to the TM mode of the anticlockwise WGM orbit~\cite{S7}. In this case, the strongly driven WGM $a_2$ can be treated classically as a number $\alpha_2\equiv \langle a_2 \rangle= 2E_2/\kappa_2$, with $\kappa_2$ the linewidth (FWHM) of the mode, and $N_2=|\alpha_2|^2$ is the intra-cavity photon number. The linearized Hamiltonian in the interaction picture can then be obtained
\begin{equation}
H^{\rm St.}_{\rm int}/\hbar= G_1 \Big( a_1^{\dag} m^{\dag} e^{-i (\Delta_1 + \omega_m)t} + a_1 m  e^{i (\Delta_1 + \omega_m)t} \Big),
\end{equation}
where $G_1=G_0 \alpha_2$ is the effective coupling rate. Since we consider a resonant pump, $\Delta_2=0$ and thus $\Delta_1=-\omega_m$, the above Hamiltonian reduces to 
\begin{equation}
H^{\rm St.}_{\rm int}= \hbar G_1 \big( a_1^{\dag} m^{\dag}  + a_1 m \big),
\end{equation}
which accounts for the two-mode squeezing interaction between the optical mode $a_1$ and magnon mode $m$, and can be used to prepare optomagnonic entangled states. This corresponds to the Stokes scattering process, where pump (TM polarized) photons convert into lower-frequency sideband (TE polarized) photons by creating magnon excitations. The corresponding quantum Langevin equations (QLEs), when taking into account the dissipation and input noise of each mode, are given by
\begin{equation}\label{Stok}
\begin{split}
 \dot{a}_1 &= -\frac{\kappa_1}{2}  a_1 - i G_1  m^{\dag} +\!\sqrt{ \kappa_1} a_1^{\rm in} ,  \\
 \dot{m} &= - \frac{\kappa_m}{2}  m - i G_1  a_1^{\dag} + \! \sqrt{\kappa_m} m^{\rm in},
\end{split}
\end{equation}
with $\kappa_1$ ($\kappa_m$) being the linewidth and $a_1^{\rm in}$ ($m^{\rm in}$) the input noise of the WGM $a_1$ (magnon mode). Note that, for simplicity, we assume the intrinsic decay rate of each WGM $\kappa_j^{i} \ll \kappa_j^e \simeq \kappa_j$ ($j=1,2$), such that for each WGM we can approximately write a single input noise operator associated with the total decay rate $\kappa_j$.

Similarly, when mode $a_1$ is resonantly pumped by a strong field (i.e., $\Delta_1=0$ and $\Delta_2=\omega_m$, cf. Fig.~\ref{fig1}(c)), e.g., by coupling the laser field to the TE mode of the anticlockwise WGM orbit~\cite{S7}, we obtain the linearized Hamiltonian in the interaction picture 
\begin{equation}\label{Haha}
H^{\rm A. S.}_{\rm int}= \hbar G_2 \big( a_2 m^{\dag}  + a_2^{\dag} m \big),
\end{equation}
where $G_2=G_0 \alpha_1$ and $\alpha_1= 2E_1/\kappa_1$. This is the state-swap interaction between the optical mode $a_2$ and magnon mode $m$, and can be used to read out the magnon state by measuring the created anti-Stokes field $a_2$. This anti-Stokes scattering corresponds to the process where pump (TE polarized) photons convert into higher-frequency anti-Stokes (TM polarized) photons by annihilating magnons. The Hamiltonian~\eqref{Haha} leads to the following QLEs:
\begin{equation}\label{anStok}
\begin{split}
 \dot{a}_2 &= -\frac{\kappa_2}{2}  a_2 - i G_2  m +\!\sqrt{ \kappa_2} a_2^{\rm in} ,  \\
 \dot{m} &= - \frac{\kappa_m}{2}  m - i G_2  a_2 + \! \sqrt{\kappa_m} m^{\rm in},
\end{split}
\end{equation}
where $a_2^{\rm in}$ is the input noise entering the WGM $a_2$.

\subsection{Mechanical motion-induced light scattering in optomechanics}

\begin{figure}[t]
\hskip0.0cm\includegraphics[width=0.85\linewidth]{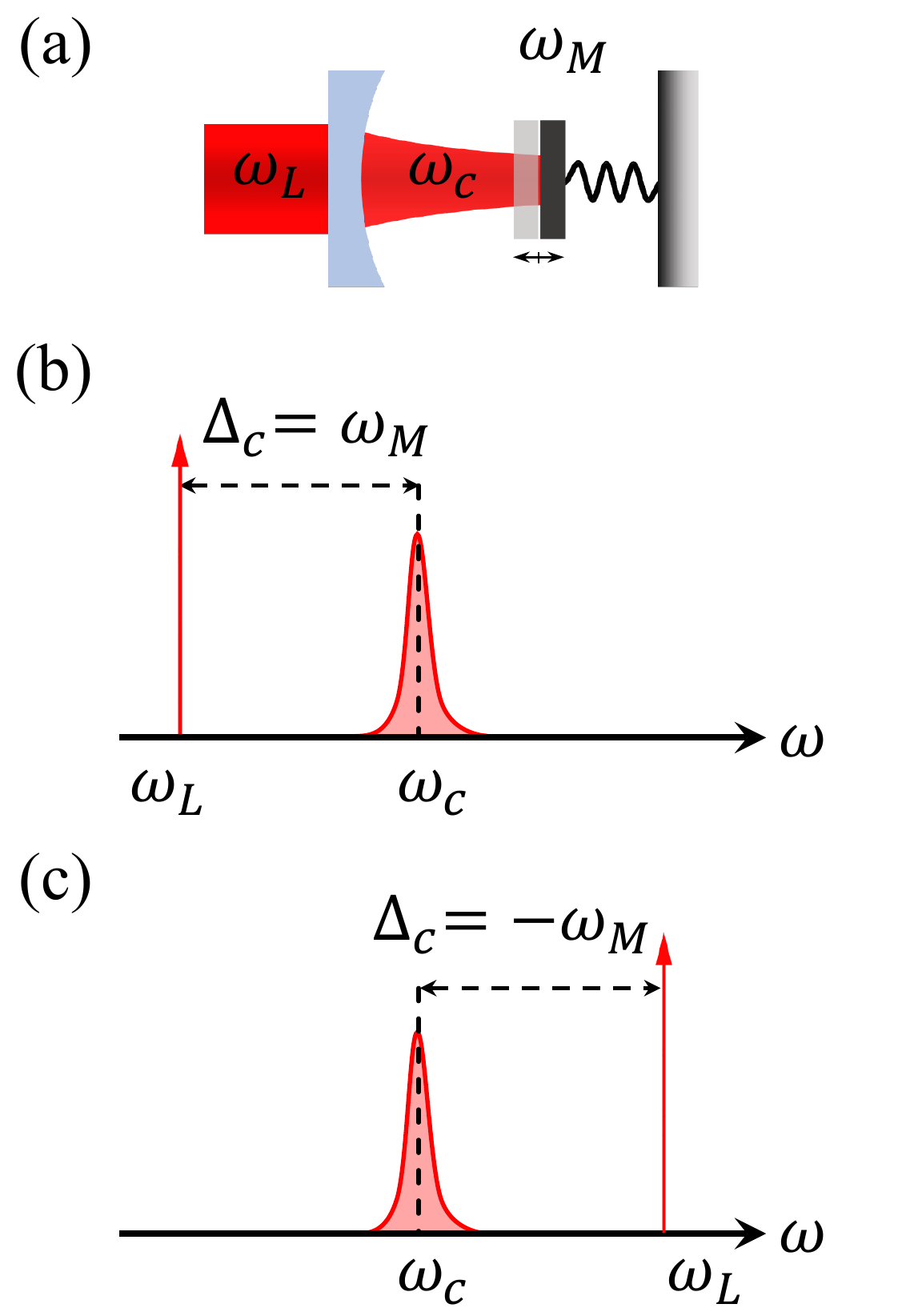} 
\caption{(a) A cavity optomechanical system: a cavity mode couples to a mechanical resonator via radiation pressure. (b) A red-detuned drive field to activate the optomechanical anti-Stokes process for realizing state swap. (c) A blue-detuned drive field to activate the optomechanical Stokes process for realizing two-mode squeezing.}
\label{fig2}
\end{figure}

The cavity optomechanical system consists of an optical cavity mode and a mechanical resonator that are coupled by the radiation pressure~\cite{OMrmp}, as depicted in Fig.~\ref{fig2}(a). The cavity is driven by a laser field that is tuned on either the red or the blue mechanical sideband for realizing the optomechanical state-swap or two-mode squeezing interaction. In particular, the former state-swap interaction will be utilized in the present protocol for building a magnon-phonon quantum network. We now show how to realize these operations and derive their effective interaction Hamiltonians. The Hamiltonian of a typical optomechanical system reads
\begin{equation}\label{Hami2}
\begin{split}
H/\hbar = \omega_c c^{\dag} c + \omega_M b^{\dag} b - g c^{\dag} c (b\,{+}\,b^{\dag})+i  \epsilon ( c^{\dag} e^{-i \omega_L t} {-} c e^{i \omega_L t}),
\end{split}
\end{equation}
where $c$ and $b$ ($c^{\dag}$ and $b^{\dag}$) are the annihilation (creation) operators of the cavity and mechanical modes, respectively, $\omega_c$ and $\omega_M$ are their resonance frequencies, $g$ is the single-photon optomechanical coupling rate, and $\epsilon=\!\sqrt{P_L \kappa_c^{e}/\hbar \omega_L}$ is the coupling strength between the cavity (with external decay rate $\kappa_c^{e}$) and the laser field (with frequency $\omega_L$ and power $P_L$). This Hamiltonian leads to the following QLEs, by including the dissipations and input noises of the two modes, in the frame rotating at the laser frequency
\begin{equation}
\begin{split}
\dot{c} &= -\Big( \frac{\kappa_c}{2} + i \Delta_c \Big) c + i g c (b + b^{\dag}) + \epsilon +\!\sqrt{\kappa_c} c^{\rm in} ,  \\
\dot{b} &= -\Big( \frac{\gamma}{2} + i \omega_M \Big) b + i g c^{\dag} c + \!\sqrt{\gamma} b^{\rm in},
\end{split}
\end{equation}
where $\Delta_c=\omega_c - \omega_L$ is the cavity-laser detuning, $\kappa_c$ ($\gamma$) is the linewidth and $c^{\rm in}$ ($b^{\rm in}$) is the input noise of the cavity (mechanical) mode. When the cavity is strongly driven, the cavity field amplitude $|\langle c \rangle| \gg 1$, which allows us to linearize the system dynamics around the classical averages (by writing the operators as $c=\langle c \rangle + \delta c$ and $b=\langle b \rangle + \delta b$) and obtain the linearized QLEs for the quantum fluctuations
\begin{equation}\label{QLEsss}
\begin{split}
\delta \dot{c} &= -\Big( \frac{\kappa_c}{2} + i \tilde{\Delta}_c \Big) \delta c + i G (\delta b + \delta b^{\dag} ) +\!\sqrt{ \kappa_c} c^{\rm in} ,  \\
\delta \dot{b} &= -\Big( \frac{\gamma}{2} + i \omega_M \Big) \delta b + i G (\delta c + \delta c^{\dag}) +\! \sqrt{\gamma} b^{\rm in},
\end{split}
\end{equation}
with $\tilde{\Delta}_c =\Delta_c -2g {\rm Re} \langle b \rangle$ being the effective detuning, which includes the frequency shift caused by the optomechanical interaction, and $G=g \langle c \rangle$ the effective coupling rate, where $\langle c \rangle=\frac{\epsilon}{(\kappa_c/2) +i \tilde{\Delta}_c}$. Note that in getting QLEs~\eqref{QLEsss}, we choose a phase reference such that $\langle c \rangle$ (and thus $G$) is real, and the expression of $\langle c \rangle$ is the steady-state solution.

To see more clearly how to operate the system to realize the two basic optomechanical interactions, we move to another interaction picture by introducing the slowly moving operators $\delta \tilde c(t) =\delta c(t) e^{i \tilde{\Delta}_c t}$ and $\delta \tilde b(t) =\delta b(t) e^{i \omega_M t}$. The QLEs then take the following form in the interaction picture (for simplicity we remove the tilde signs on the operators):
\begin{equation}
\begin{split}
\delta \dot{c} &= {-}\frac{\kappa_c}{2} \delta c + i G \Big[ \delta b\, e^{i (\tilde{\Delta}_c -\omega_M) t} {+} \delta b^{\dag} e^{i (\tilde{\Delta}_c +\omega_M) t} \Big] +\!\sqrt{ \kappa_c} c^{\rm in} ,  \\
\delta \dot{b} &= {-}\frac{\gamma}{2} \delta b + i G \Big[ \delta c\, e^{i (\omega_M -\tilde{\Delta}_c) t}  {+} \delta c^{\dag} e^{i (\omega_M +\tilde{\Delta}_c) t} \Big] + \! \sqrt{\gamma} b^{\rm in} .
\end{split}
\end{equation}
For a red-detuned driving field, $\tilde{\Delta}_c = \omega_M$ (cf. Fig.~\ref{fig2}(b)), we obtain the approximate equations
\begin{equation}\label{RedQLE}
\begin{split}
\delta \dot{c} &\approx -\frac{\kappa_c}{2} \delta c + i G \delta b +\!\sqrt{ \kappa_c} c^{\rm in} ,  \\
\delta \dot{b} &\approx - \frac{\gamma}{2} \delta b + i G \delta c + \! \sqrt{\gamma} b^{\rm in},
\end{split}
\end{equation}
which corresponds to the linearized interaction Hamiltonian
\begin{equation}\label{BSHami}
H^{\rm A. S.}_{\rm int} = -\hbar G (c^{\dag} b + c b^{\dag}),
\end{equation}
and the process of anti-Stokes scattering for realizing the state-swap interaction between the mechanics and the cavity field. While for a blue-detuned driving field, $\tilde{\Delta}_c = -\omega_M$ (cf. Fig.~\ref{fig2}(c)), we get 
\begin{equation}\label{BlueQLE}
\begin{split}
\delta \dot{c} &\approx -\frac{\kappa_c}{2} \delta c + i G \delta b^{\dag} +\!\sqrt{ \kappa_c} c^{\rm in} ,  \\
\delta \dot{b} &\approx - \frac{\gamma}{2} \delta b + i G \delta c^{\dag} + \! \sqrt{\gamma} b^{\rm in},
\end{split}
\end{equation}
which corresponds to the interaction Hamiltonian 
\begin{equation}\label{TMSHami}
H^{\rm St.}_{\rm int} = -\hbar G (c^{\dag} b^{\dag} + c b),
\end{equation}
and the process of Stokes scattering for achieving the two-mode squeezing interaction between the mechanical and cavity modes. Note that in deriving Eqs.~\eqref{RedQLE} and \eqref{BlueQLE} we assume $\gamma,\, \kappa_c, \, G \ll \omega_M$, such that the non-resonant fast oscillating terms play a negligible role and can be neglected.

\begin{figure*}[t]
\centering
\hskip-0.5cm\includegraphics[width=0.85\linewidth]{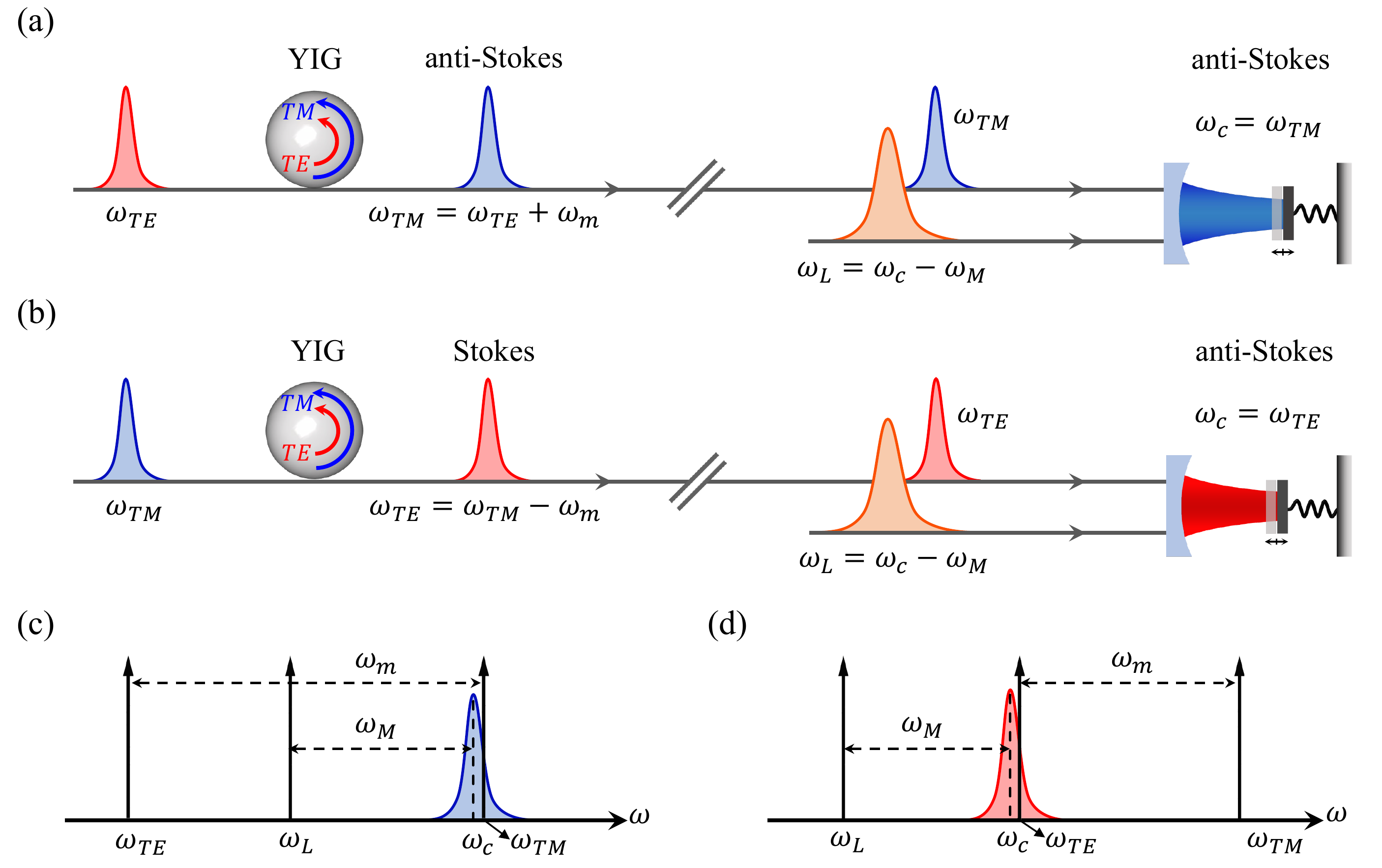} 
\caption{Sketch of the distant magnon-to-phonon state transfer protocol (a) and the corresponding mode frequencies (c). The nonlocal magnon-phonon entanglement protocol (b) and the associated mode frequencies (d). }
\label{fig3}
\end{figure*}

\section{Distant Magnon-to-phonon state transfer}
\label{transf}

Equipped with all the necessary tools, we now proceed to describe our protocol. Note that the cavity optomagnonic system using a YIG sphere currently works in the weak coupling regime, where the enhanced optomagnonic coupling rate (potentially in MHz) is much smaller than the decay rate of the WGM (from $10^2$ MHz to GHz), $G_j \ll \kappa_j$~\cite{S7,S8,S9}. The large cavity decay disables many quantum protocols that require the cooperativity $\CC=G_j^2/(\kappa_j \kappa_m) >1$, considering also $\kappa_m/2\pi$ is hardly below 1 MHz for YIG~\cite{S1,S2,S3}. However,  this deficiency of the system can be evaded by using fast optical pulses~\cite{Simon16,Simon18,Jie21,Xie,Heqy}. 

In the first part, we aim to realize a distant magnon-to-phonon (quantum) state transfer. Specifically, as shown in Fig.~\ref{fig3}(a), by using laser pulses to successively activate the anti-Stokes scatterings in the optomagnonic and optomechanical systems, an arbitrary magnonic state can be transferred to a mechanical resonator that can have a much longer coherence time. We have introduced the interaction Hamiltonian~\eqref{anStok} of the optomagnonic anti-Stokes scattering in Sec.~\ref{bas} A,  associated with the process where a TE polarized pulse couples to a WGM and generates TM polarized anti-Stokes photons in another WGM by annihilating magnons. We consider laser pulses with duration much shorter than the magnon lifetime~\cite{note}. In this case, the dissipation of the magnon mode within the pulse duration is negligibly small, and we thus neglect it for simplicity. This leads to the following QLEs during the pulse interaction:
\begin{equation}\label{sta}
\begin{split}
 \dot{a}_2 &= -\frac{\kappa_2}{2}  a_2 - i G_2  m +\!\sqrt{ \kappa_2} a_2^{\rm in} ,  \\
 \dot{m} &=  - i G_2  a_2 .
\end{split}
\end{equation}
To simplify the model, we consider a flattop pulse and thus a constant coupling $G_2$ during the pulse. Given also the fact that a weak coupling $G_2 \ll \kappa_2$, one can then adiabatically eliminate the cavity, and obtain $ a_2 \simeq \frac{2}{\kappa_2} \big( -i G_2  m +\!\sqrt{\kappa_2} a_2^{\rm in} \big)$. By using the input-output relation $ a_2^{\rm out}\,{=}\,\sqrt{\kappa_2}  a_2 - a_2^{\rm in} $~\cite{collett}, we obtain 
\begin{equation}\label{a2out-m}
\begin{split}
 a_2^{\rm out} &= - i \sqrt{2 \GG_2} m + a_2^{\rm in},  \\
 \dot{m} &= -\GG_2  m - i \sqrt{2 \GG_2} a_2^{\rm in}, 
\end{split}
\end{equation}
where $\GG_2  \equiv 2 G_2^2/\kappa_2$. Following~\cite{Hofer}, we define a set of normalized temporal modes for the WGM driven by a pulse of duration $\tau_2$ 
\begin{equation}\label{tempo}
\begin{split}
 A_2^{\rm in} (\tau_2) &= \sqrt{  \frac{2\GG_2}{e^{2\GG_2 \tau_2} -1} } \int_0^{\tau_2} e^{\GG_2 s} a_2^{\rm in} (s)\, ds , \\
 A_2^{\rm out} (\tau_2) &= \sqrt{ \frac{ 2\GG_2}{ 1- e^{-2\GG_2 \tau_2} } } \int_0^{\tau_2} e^{- \GG_2 s} a_2^{\rm out} (s)\, ds ,
 \end{split}
\end{equation}
which satisfy the canonical commutation relation $[A^j, \,\, A^{j \dag}] \,{=}\, 1, j \,{=}\, \{ {\rm in}, \,\, {\rm out}\} $. Therefore, by integrating~\eqref{a2out-m} we obtain the following solutions (see Appendix):
\begin{equation}\label{Aout-m}
\begin{split}
 A_2^{\rm out} (\tau_2) &= - i \sqrt{ 1- e^{-2\GG_2 \tau_2} } m(0) + e^{-\GG_2 \tau_2} A_2^{\rm in} (\tau_2) ,  \\
 m (\tau_2) &= e^{-\GG_2 \tau_2} m(0) - i \sqrt{ 1- e^{-2\GG_2 \tau_2} } A_2^{\rm in}(\tau_2). 
 \end{split}
\end{equation}
From these solutions, we can extract a propagator $L_2(\tau_2)$ that satisfies $A_2^{\rm out} (\tau_2) = L_2^{\dag} (\tau_2) \, A_2^{\rm in} (\tau_2) \, L_2(\tau_2)$ and $m(\tau_2) = L_2^{\dag} (\tau_2) \, m(0) \, L_2(\tau_2)$, given by~\cite{Xie}
 \begin{equation}\label{end}
 L_2(\tau_2) = e^{-i \sqrt{S'}  A_2^{\rm in \dag} m }  e^{\GG_2 \tau_2 (A_2^{\rm in \dag} A_2^{\rm in} - m^{\dag} m) }  e^{i \sqrt{S'} A_2^{\rm in} m^{\dag} }, 
 \end{equation}
where $S'=S  e^{2\GG_2 \tau_2}$, with $S=1-e^{-2\GG_2 \tau_2}$ ($0<S<1$) being the optomagnonic state conversion efficiency, which depends on the pulse strength and duration. It is clear that when $\GG_2 \tau_2 \gg 1$, $S\to 1$, and we thus get $A_2^{\rm out} (\tau_2) \simeq - i  m(0)$, which implies that the magnon state is perfectly transferred to the optical mode, apart from a phase difference $-i$. A similar mechanism by using the light anti-Stokes scattering has been adopted to read out the mechanical state in optomechanics~\cite{Hofer,tjk,jie18,Simon16,Simon18}. 

We assume that the optomagnonic system is in an initial state 
 \begin{equation}
\rho_{m,2}(0)= \sum_{n,s=0}^{\infty} c_{n,s} |n\rangle \langle s|_m \otimes |0\rangle \langle 0|_{2},
 \end{equation}
before sending the TE polarized pulse to couple to the WGM. To be generic, the magnon mode is assumed in an {\it arbitrary} state, which can be expanded in the Fock-state basis with arbitrary real coefficients $c_{n,s} \ge 0$, i.e., $\rho_m(0)= \sum_{n,s=0}^{\infty} c_{n,s} |n\rangle \langle s|_m$~\cite{jie10}, and the WGM $a_2$ in vacuum state. As discussed, the TE polarized pulse (with duration $\tau_2$ and strength yielding an effective coupling $G_2$) activates the anti-Stokes BLS, associated with the time-evolution propagator $L_2(\tau_2)$. The system, at the end of the pulse, evolves to be
 \begin{equation}
 \rho_{m,2}(\tau_2)=\sum_{n,s=0}^{\infty} c_{n,s} (-i)^n i^s S^{\frac{n+s}{2}}  |0\rangle \langle 0|_{m} \otimes  |n\rangle \langle s|_2. 
 \end{equation}
Clearly, the initial magnon state $\rho_m(0)$ is transferred to the TM polarized pulse (anti-Stokes photons) in the state $\rho_{\rm TM}(\tau_2)=\sum_{n,s=0}^{\infty} c_{n,s} (-i)^n i^s S^{\frac{n+s}{2}} |n\rangle \langle s|_2$. The fidelity in this state transfer is reduced owing to a nonunity conversion efficiency $S<1$ (apart from a phase difference $(-i)^n i^s$). 

The TM polarized pulse then transmits, through a fiber, to a distant optomechanical system and {\it resonantly} drives the optomechanical cavity, see Figs.~\ref{fig3}(a), (c). Meantime, the cavity is driven by another red-detuned pulse, which activates the optomechanical anti-Stokes process for realizing the state-swap between the cavity field and the mechanics. Such a configuration has been employed to transfer the state of an electromagnetic field to a mechanical resonator~\cite{Jie19,JieQST,jie18,zoller,tian}.  Owing to the similarity between the optomagnonic and optomechanical anti-Stokes scatterings (in their interaction Hamiltonians \eqref{Haha} and \eqref{BSHami}), following the same approach as from Eq.~\eqref{sta} to Eq.~\eqref{end}, we can extract a propagator associated with the pulse (with duration $\tau_b$) that activates the optomechanical anti-Stokes process
\begin{equation}\label{UoM}
L_{\rm oM}(\tau_b) = e^{i \sqrt{W'} C^{\rm in} b^{\dag} }  e^{\GG \tau_b ( b^{\dag} b - C^{\rm in \dag} C^{\rm in} ) }  e^{-i \sqrt{W'} C^{\rm in \dag} b }, 
\end{equation}
where $C^{\rm in}$ denotes the temporal mode of the input field (entering the optomechanical cavity), $\GG  \equiv 2 G^2/\kappa_c$, and $W'=W  e^{2\GG \tau_b}$, with $W=1-e^{-2\GG \tau_b}$ being the optomechanical state conversion efficiency, $0<W<1$. Similarly, we assume the pulse duration to be much shorter than the mechanical lifetime and a weak coupling $G \ll \kappa_c$, which can be easily satisfied because of a relatively longer mechanical lifetime~\cite{OMrmp}.

We further assume that the mechanical mode is prepared in the ground state $|0\rangle_M$ (for a GHz resonator requiring a bath temperature of tens of mK~\cite{Simon16,Simon18}). Alternatively, the mechanical resonator can also be precooled to its ground state using a red-detuned light~\cite{chen}. The previously generated TM polarized pulse now acts as the input field into the optomechanical cavity, and the evolution of the system can be solved by applying the propagator $L_{\rm oM}(\tau_b)$ onto the initial state~\cite{notee}
\begin{equation}
\rho_{\rm oM}(0)=\sum_{n,s=0}^{\infty} c_{n,s} (-i)^n i^s S^{\frac{n+s}{2}}  |n\rangle \langle s|_o \otimes  |0\rangle \langle 0|_{M}. 
\end{equation}
At the end of the pulse, we obtain the state $\rho_{\rm oM}(\tau_b)= L_{\rm oM}(\tau_b) \rho_{\rm oM}(0) L_{\rm oM}^{\dag}(\tau_b)$, given by
\begin{equation}
\rho_{\rm oM}(\tau_b)=\sum_{n,s=0}^{\infty} c_{n,s} (SW)^{\frac{n+s}{2}}   |0\rangle \langle 0|_{o}  \otimes |n\rangle \langle s|_M.
\end{equation}
It shows that the state of the TM polarized pulse $\rho_{\rm TM}(\tau_2)$ is transferred to the mechanical resonator, which is in the state $\rho_{M}(\tau_b)=\sum_{n,s=0}^{\infty} c_{n,s} (SW)^{\frac{n+s}{2}}  |n\rangle \langle s|_M$, with a reduced fidelity due to a nonunity conversion efficiency $W<1$. 

Looking at the whole process, after two state-swap operations (magnon-to-photon and photon-to-phonon), an arbitrary magnon state $\rho_m(0)$ is successfully transferred to the mechanical mode in the state $\rho_{M}(\tau_b)$. The fidelity is determined by the product of the two conversion efficiencies $S$ and $W$ in the two anti-Stokes processes.  We provide two concrete examples: When the magnon mode is initially in a Fock state $|\phi \rangle_m = |n\rangle$, we obtain the final transferred mechanical state $|\phi' \rangle_M = (SW)^{\frac{n}{2}} |n\rangle$. The fidelity is $\FF=| \langle  \phi | \phi' \rangle |^2 = (SW)^n$, which reduces with an increasing $n$. For an initial superposition state $|\phi \rangle_m =\frac{1}{\sqrt{2}} ( |0\rangle + |1\rangle)$, we get $|\phi' \rangle_M =\frac{1}{\sqrt{2}} \big( |0\rangle + \sqrt{SW} |1\rangle \big)$, and the fidelity $\FF= \frac{1}{4}\big(1+\sqrt{SW}\big)^2$. For perfect state conversions $S,W \to 1$, so the fidelity $\FF \to 1$. Note that, because of the nonunity state conversion efficiencies in actual experiments, the experiment will be repeated many times using a sequence of pulses. The fidelity then can be interpreted as the probability of the successful/perfect magnon-to-phonon state transfer in a single experiment run~\cite{Simon2}. The mechanical state can be stored within the mechanical coherence time and retrieved by sending a weak red-detuned read pulse to the cavity and measuring the cavity output field~\cite{Simon,Simon2,Simon16,Simon18}. 


It should be noted that, for simplifying the calculations, we adopt laser pulses with duration much shorter than the magnon lifetime $\sim$ 1 $\mu$s for the optomagnonic system. This allows us to neglect the loss of the magnons. However, the small pulse duration also reduces the conversion efficiency and thus the fidelity in the state transfer. This can be compensated by increasing the pulse strength or the single-photon coupling rate $G_0$, e.g., by reducing the mode volumes and increasing the mode overlap of the magnon and optical fields, purifying and doping YIG~\cite{Greek}, coupling WGMs to a magnetic vortex~\cite{SVK18}, and utilizing the epsilon-near-zero medium~\cite{SVK21}, etc. Let us estimate the optomagnonic conversion efficiency using promising parameters. Taking a cavity decay rate $\kappa_2/2\pi \, {\sim}\, 500$ MHz, an effective coupling $G_2/2\pi \, {\sim} \,10$ MHz~\cite{S8,Flat}, and a pulse duration $\tau_2=40$ ns, we obtain $\GG_2 \tau_2 = \frac{2 G_2^2 }{\kappa_2} \tau_2 \simeq 0.10$, and a moderate efficiency $S=1-e^{-2 \GG_2 \tau_2} \simeq 0.18$.

For the mechanical system, because of its much longer lifetime, longer pulses could be used to increase the conversion efficiency. We adopt the parameters from an optomechanical experiment~\cite{Simon16}: a mechanical mode of frequency $\omega_M/2\pi \,\,{=}\,5.3$ GHz and damping rate $\gamma/2\pi \,\,{=}\, 4.8$ kHz (corresponding to lifetime $2\pi/\gamma \,\,{\simeq}\,\, 0.2$ ms), a cavity decay rate $\kappa_c/2\pi \,\,{=}\, 1.3$ GHz, a pulse duration $\tau_b \,{=}\, 55$ ns, and an effective coupling $G/2\pi \,\,{=}\, 50$ MHz. These parameters yield $\GG \tau_b \,{=}\, \frac{2 G^2 }{\kappa_c} \tau_b \,{\simeq}\, 1.33$, and a high conversion efficiency $W\,{=}\,1-e^{-2 \GG \tau_b} \simeq 0.93$. Note that there are many different types of optomechanical devices (see Fig.7 and Table II in \cite{OMrmp} for their characteristic parameters), among which a high conversion efficiency together with a long mechanical lifetime is preferred for our protocol.

\begin{figure}[t]
\centering
\hskip-0cm\includegraphics[width=0.85\linewidth]{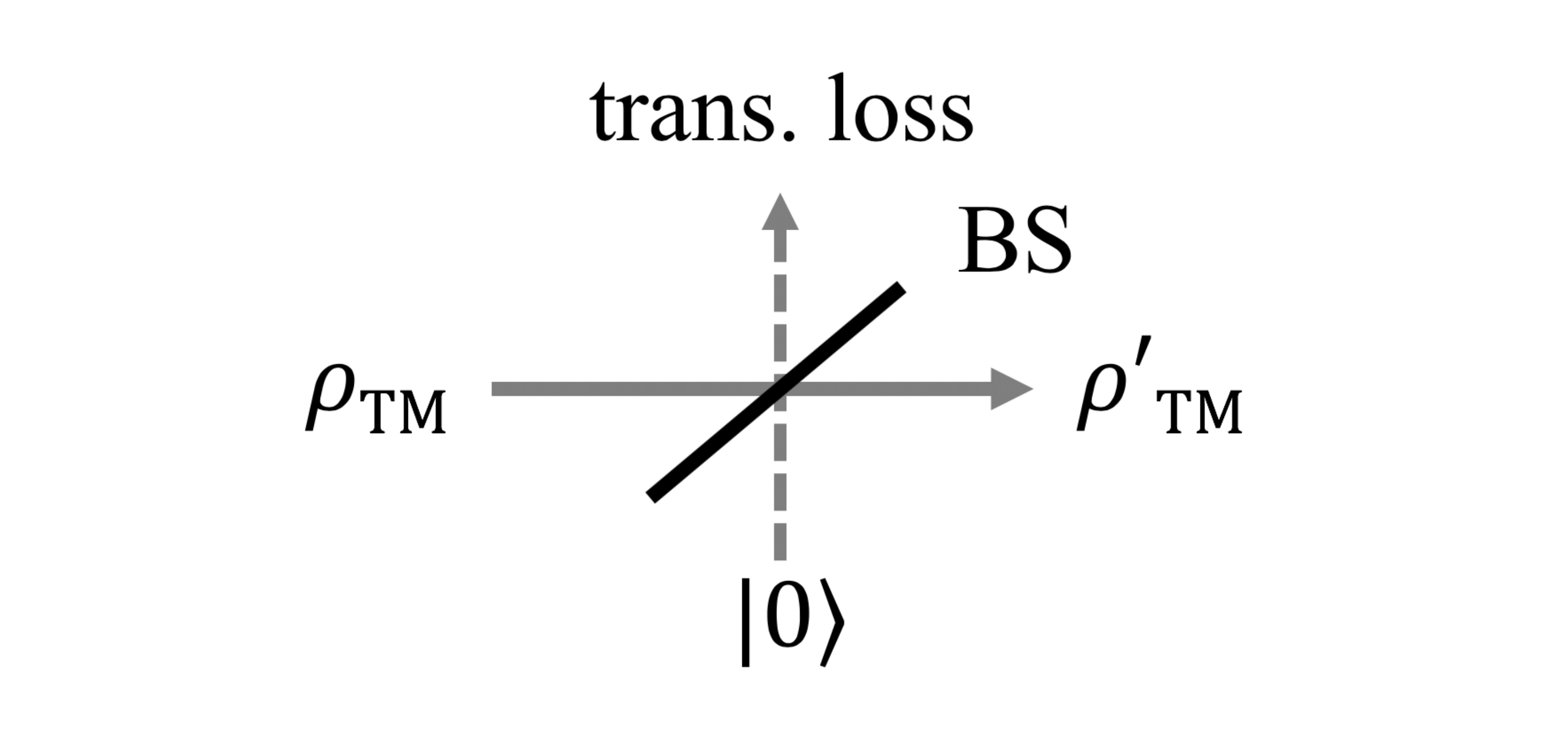} 
\caption{The transmission loss modeled by a beamsplitter (BS), of which the reflection denotes the loss and the transmission represents the pulse after suffering the loss.}
\label{fig4}
\end{figure}

\subsection{Effect of the linear loss in pulse transmission}

In the preceding section, for simplicity, we neglect the transmission loss of the pulse from the magnonic to the mechanical system. However, for building a long-distance quantum network, this loss can be significant. The transmission loss is linear with the transmission distance, and thus can be modeled by a linear beamsplitter~\cite{ULeo}, as depicted in Fig.~\ref{fig4}. The pulse enters one input port of the beamsplitter with the other input in vacuum, of which the reflected part denotes the loss and the transmitted part represents the pulse after suffering the loss.

Including the loss in the transmission time $\Delta t$, the generated TM pulse state $\rho_{\rm TM}(\tau_2)$ (for an arbitrary magnon state) turns into the following state when reaching the optomechanical cavity~\cite{jie10}:
\begin{equation}
\begin{split}
 \rho_{\rm TM}&(\tau_2 + \Delta t)=\sum_{n,s=0}^{\infty} c_{n,s} (-i)^n i^s S^{\frac{n+s}{2}} \times  \\ 
 &\,\, \sum_{m=0}^{{\rm min}(n,s)} \!\! \sqrt{\frac{n!s!}{(m!)^2 (n-m)!(s-m)!}} R^m T^{\frac{n+s}{2}-m}  | n{-}m\rangle \langle s{-}m |_2 ,
 \end{split}
 \end{equation}
where $R$ ($T$) is the reflectance (transmittance) of the beamsplitter, and we assume a lossless beamsplitter $R+T=1$. Then the initial state of the optomechanical system, before activating the optomechanical anti-Stokes process, is~\cite{notee}
\begin{equation}
\rho'_{\rm oM}(0)=  \rho_{\rm TM}(\tau_2 + \Delta t) \otimes  |0\rangle \langle 0|_{M}.
\end{equation}
We obtain the state $\rho'_{\rm oM}(\tau_b)= L_{\rm oM}(\tau_b) \rho'_{\rm oM}(0) L_{\rm oM}^{\dag}(\tau_b)$ soon after the state-swap interaction, which is
\begin{equation}\label{RRR}
\begin{split}
\rho'_{\rm oM}(\tau_b)&=\sum_{n,s=0}^{\infty} c_{n,s} S^{\frac{n+s}{2}}  \sum_{m=0}^{{\rm min}(n,s)} \!\! \sqrt{\frac{n!s!}{(m!)^2 (n-m)!(s-m)!}}   \\
& \times R^m (T W)^{\frac{n+s}{2}-m}  |0\rangle \langle 0|_{o}  \otimes |n-m \rangle \langle s-m |_M.
 \end{split}
\end{equation}
The above state does not look intuitive. Let us consider some specific cases to see the physics more clearly. For an initial magnon Fock state $|\phi \rangle_m = |n\rangle$, Eq.~\eqref{RRR} yields the following transferred mechanical state after the transmission loss:
\begin{equation}
\begin{split}
\rho'_M(\tau_b)&= S^{n}  \sum_{m=0}^{n}\frac{n!}{m! (n-m)!} R^m (T W)^{n-m}  |n{-}m \rangle \langle n{-}m |_M.
 \end{split}
\end{equation}
More specifically, for $n=1$, i.e., an initial single-magnon state $| 1 \rangle_m$, we get 
\begin{equation}
\begin{split}
\rho'_M(\tau_b)= S \Big( T W | 1 \rangle \langle 1 |_M + R | 0 \rangle \langle 0 |_M   \Big).
 \end{split}
\end{equation}
The fidelity between the initial magnon state and the transferred mechanical state is thus
\begin{equation}
\FF= \langle  \phi | \rho'_M(\tau_b) | \phi \rangle  = STW,
\end{equation}
which agrees with our previous result for a magnon Fock state $\FF \,{=}\,\, (SW)^n$ ($n\,\,{=}\,\,1$) for a lossless transmission $T\,\,{=}\,\,1$. Apparently, the fidelity is proportional to the two conversion efficiencies and the proportion of the transmitted pulse after the loss. Considering fiber loss of 0.2 dB/km at telecom wavelengths, for a distance of 1 km (10 km), the total fiber loss is 0.2 dB (2 dB), corresponding to $T\simeq 0.955$ (0.631). Taking the two realistic conversion efficiencies $S\,\,{=}\,\,0.18$ and $W\,\,{=}\,\,0.93$ estimated in the preceding section, we obtain fidelity $\FF \,\,{=}\,\, STW \,\,{=}\,\, 0.16$ (0.10) for transferring a single-magnon state $ |1\rangle_m$ to a long-lived mechanical mode. This means that for 100 runs of the experiment, about 16 (10) times the mechanical mode is in the single-phonon state $ |1\rangle_M$.

For the initial superposition state $|\phi \rangle_m =\frac{1}{\sqrt{2}} ( |0\rangle + |1\rangle)$, using the result of Eq.~\eqref{RRR}, we obtain the mechanical state
\begin{equation}
\begin{split}
\rho'_M(\tau_b)= \frac{1}{2} & \Big[ \big(1+ S R \big) | 0 \rangle \langle 0 |_M +\! \sqrt{STW}  |0 \rangle \langle 1|_M  \\
&+ \! \sqrt{STW}  |1 \rangle \langle 0|_M +STW |1 \rangle \langle 1|_M   \Big].
 \end{split}
\end{equation}
The fidelity is then
\begin{equation}
\FF =  \frac{1}{4} \Big( 1+SR + 2\sqrt{STW} +STW  \Big ) ,
\end{equation}
and in the loss-free limit $T \to 1$ ($R \to 0$), it becomes $\FF= \frac{1}{4}\big(1+\sqrt{SW}\big)^2$, which agrees with the result in the preceding section for a lossless transmission.

\section{Nonlocal macroscopic Magnon-phonon entanglement}
\label{entang}

Under certain circumstances, the quantum network requires its nodes to be entangled~\cite{Kimble,Wehner} and many quantum protocols, like quantum repeaters~\cite{Gisin} and teleportation~\cite{Simon2}, require quantum entanglement.  In this section, we show that our system also allows to prepare a nonlocal macroscopic magnon-phonon entangled state using optical pulses. Specifically, laser pulses are sent to successively activate the optomagnonic Stokes scattering and the optomechanical anti-Stokes scattering, see Figs.~\ref{fig3}(b), (d). The former generates an entangled state of the magnons in YIG and the Stokes photons in the pulse, and the latter maps the pulse state to the mechanical resonator, and thus a nonlocal magnon-phonon entangled state is established. The maximum distance between the two entangled subsystems is determined by the relatively short magnon lifetime, beyond which the magnon state degrades and the entanglement dies out.

We introduce the interaction Hamiltonian~\eqref{Stok} accounting for the optomagnonic Stokes scattering. As before, we neglect the magnon dissipation during the short pulse, and obtain the QLEs
\begin{equation}
\begin{split}
 \dot{a}_1 &= -\frac{\kappa_1}{2}  a_1 - i G_1  m^{\dag} +\!\sqrt{ \kappa_1} a_1^{\rm in} ,  \\
 \dot{m} &=  - i G_1  a_1^{\dag} .
\end{split}
\end{equation}
By adiabatically eliminating the cavity field, we get $ a_1 \simeq \frac{2}{\kappa_1} \big( -i G_1  m^{\dag} +\!\sqrt{\kappa_1} a_1^{\rm in} \big)$, and using the input-output relation $ a_1^{\rm out}=\sqrt{\kappa_1}  a_1 - a_1^{\rm in} $, we obtain 
\begin{equation}\label{a1out-m}
\begin{split}
 a_1^{\rm out} &= - i \sqrt{2 \GG_1} m^{\dag} + a_1^{\rm in},  \\
 \dot{m} &= \GG_1  m - i \sqrt{2 \GG_1} a_1^{\rm in \dag}, 
\end{split}
\end{equation}
where $\GG_1  \equiv 2 G_1^2/\kappa_1$. Again, we define the normalized temporal modes for the WGM driven by a pulse of duration $\tau_1$
\begin{equation}
\begin{split}
 A_1^{\rm in} (\tau_1) &= \sqrt{  \frac{2\GG_1}{1 - e^{- 2\GG_1 \tau_1}} } \int_0^{\tau_1} e^{- \GG_1 s} a_1^{\rm in} (s)\, ds , \\
 A_1^{\rm out} (\tau_1) &= \sqrt{ \frac{ 2\GG_1}{ e^{ 2\GG_1 \tau_1} -1} } \int_0^{\tau_1} e^{ \GG_1 s} a_1^{\rm out} (s)\, ds .
 \end{split}
\end{equation}
By integrating~\eqref{a1out-m}, we obtain
\begin{equation}\label{AAAmmmt1}
\begin{split}
 A_1^{\rm out} (\tau_1) &= - i \sqrt{ e^{2 \GG_1 \tau_1} -1} m^{\dag}(0) + e^{\GG_1 \tau_1} A_1^{\rm in} (\tau_1) ,  \\
 m (\tau_1) &= e^{\GG_1 \tau_1} m(0) - i \sqrt{ e^{2 \GG_1 \tau_1} -1} A_1^{\rm in \dag}(\tau_1),
 \end{split}
\end{equation}
from which a propagator $L_1(\tau_1)$ is extracted, satisfying $A_1^{\rm out} (\tau_1) = L_1^{\dag} (\tau_1) \, A_1^{\rm in} (\tau_1) \, L_1(\tau_1)$ and $m(\tau_1) = L_1^{\dag} (\tau_1) \, m(0) \, L_1(\tau_1)$, which is~\cite{tjk}
 \begin{equation}
 L_1(\tau_1) = e^{-i \tanh r  \, A_1^{\rm in \dag} m^{\dag} }  \cosh r^ {(-1- A_1^{\rm in \dag} A_1^{\rm in} - m^{\dag} m) }  e^{i  \tanh r \, A_1^{\rm in} m }, 
 \end{equation}
where we introduce the squeezing parameter $r$ through $\cosh r = e^{\GG_1 \tau_1}$ and $\tanh r =\sqrt{1-e^{-2 \GG_1 \tau_1}}$. It is clear that the squeezing $r$ increases with the product $\GG_1 \tau_1$, i.e., the product of the pulse strength and duration. 

For an initial state $| 0 \rangle_m |0 \rangle_1$~\cite{nnnot}, the optomagnonic system is prepared, at the end of the pulse, in the state
\begin{equation}\label{sss}
| \phi \rangle_{m,1}(\tau_1) = (\cosh r)^{-1} \sum_{n=0}^{\infty} (\tanh r)^n  |n,n \rangle_{m,1},
\end{equation}
which is a two-mode squeezed vacuum state (TMSVS) of the magnon mode and the TE polarized pulse (Stokes photons). The entanglement of this state is linked to the squeezing parameter as the logarithmic negativity $E_N=2 r$. Using the parameters $\kappa_1/2\pi \, {\sim}\, 500$ MHz, $G_1/2\pi \, {\sim} \,10$ MHz~\cite{S8,Flat}, and a pulse duration $\tau_1 \,{=}\,30$ ns, we obtain $\GG_1 \tau_1 \,{\simeq}\, 0.075$, and a squeezing parameter $r \simeq 0.39$, and the entanglement $E_N \simeq 0.78$.

\begin{figure}[t]
\centering
\hskip-0.4cm\includegraphics[width=0.85\linewidth]{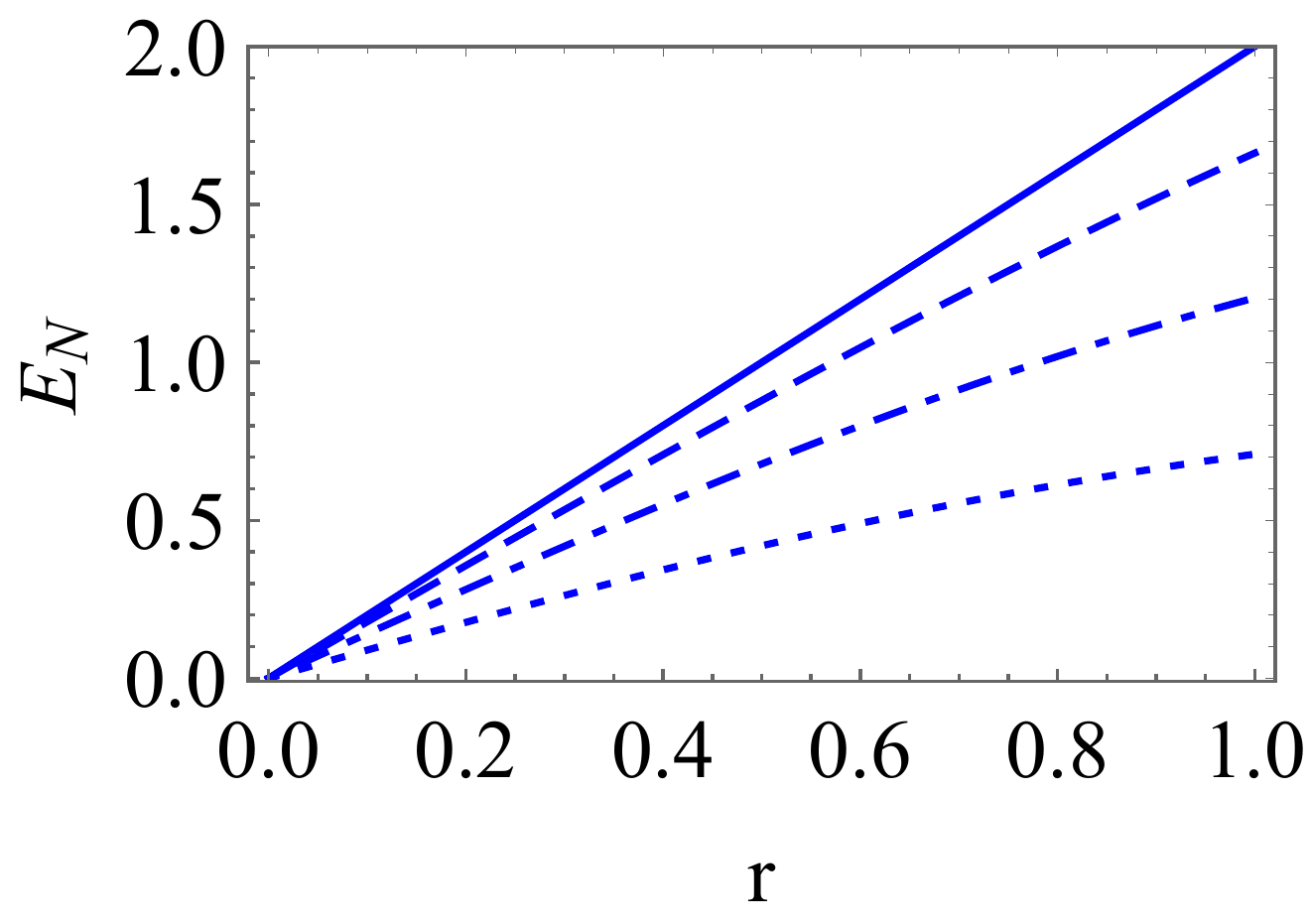} 
\caption{Magnon-phonon entanglement (logarithmic negativity $E_N$) versus the squeezing $r$ for different conversion efficiencies $W$. From upper to lower curves: $W=1$, 0.8, 0.5, and 0.2.}
\label{fig5}
\end{figure}

The TE pulse then transmits to the optomechanical system and resonantly drives the optical cavity, which is simultaneously driven by another red-detuned pulse for realizing the optomechanical state-swap operation (cf. Fig.~\ref{fig3}(d)). The corresponding propagator $L_{\rm oM}(\tau_b)$ has been introduced in Eq.~\eqref{UoM}. As in the preceding section, the mechanical resonator is prepared in the ground state. By applying the propagator $L_{\rm oM}(\tau_b)$ onto the initial state
\begin{equation}
| \phi \rangle_{m,1,M}(0) = (\cosh r)^{-1} \sum_{n=0}^{\infty} (\tanh r)^n  |n,n \rangle_{m,1} \otimes |0 \rangle_{M},
\end{equation}
we obtain the joint state $| \phi \rangle_{m,1,M}(\tau_b) = L_{\rm oM}(\tau_b) | \phi \rangle_{m,1,M}(0)$, at the end of the red-detuned pulse, given by
\begin{equation}
| \phi \rangle_{m,1,M}(\tau_b) = (\cosh r)^{-1} \sum_{n=0}^{\infty} (i \tanh r)^n  W^{\frac{n}{2}} |n,0,n \rangle_{m,1,M}.
\end{equation}
By tracing over the optical mode and writing in the density matrix form, we get 
\begin{equation}\label{zzz}
\rho_{m,M}(\tau_b) = (\cosh r)^{-2} \sum_{n=0}^{\infty} (\tanh r)^{2n} W^n  |n,n \rangle \langle n,n|_{m,M},
\end{equation}
which, in the limit of a unity conversion efficiency $W \, {\to}\, 1$ ($\GG \tau_b \gg 1$), becomes exactly the TMSVS of the magnon and mechanical modes, and thus the magnon and mechanical nodes are remotely entangled. For a realistic efficiency $W \simeq 0.93$ estimated in the preceding section, the entanglement is only slightly reduced. By comparing the states in Eqs.~\eqref{sss} and \eqref{zzz}, we see that the nonunity conversion efficiency plays a role in degrading the squeezing and thus the entanglement. This can be seen by making the replacement $\tanh r' =\tanh r \sqrt{W}$, where $r'$ is the effective squeezing including the effect of the nonunity conversion efficiency, and $r' \le r$. In Fig.~\ref{fig5}, we show the entanglement (logarithmic negativity $E_N=2r'$) versus the squeezing $r$ for different conversion efficiencies $W$.

We remark that the survival time of the entanglement is determined by the relatively short magnon lifetime $\sim 1$ $\mu$s, during which the transmission loss of the pulse is negligibly small. Therefore, in the entanglement protocol we do not include the transmission loss. The entanglement can be verified by activating the optomagnonic and optomechanical anti-Stokes processes, which transfer the magnonic and mechanical states to their respective anti-Stokes photons. By measuring the anti-Stokes photons, one can build a joint magnon-phonon covariance matrix, from which the entanglement can be verified and quantified. This approach has been used to measure the optomechanical~\cite{John1} and mechanical~\cite{John2} entanglement in optomechanics.

\section{Conclusion}\label{conc}

We present a protocol for building a quantum network based on magnonic and mechanical systems. The magnonic and mechanical nodes are connected by optical pulses. We show that the quantum network can realize (both quantum and classical) state transfer from the magnonic node to the long-lived mechanical node, taking into account the transmission loss for a long-distance network. We also show that the two systems can be prepared in a nonlocal macroscopic entangled state. All these utilize the Stokes and anti-Stokes light scatterings in optomagnonics and optomechanics, and optical pulses to transfer quantum states or distribute quantum correlations. As is known, quantum state transfer and entanglement distribution among different nodes are basic functions for a quantum network~\cite{Kimble,Rempe,cirac}. Our work shows the possibility to connect the seemingly separate research fields of optomagnonics and optomechanics, and to build a quantum network based on magnonic systems, which can be readily extended to a complex network with various nodes by coupling to other quantum systems, e.g., long-lived phonons, optical/microwave photons, and superconducting qubits~\cite{NakaRev}. We expect our work may offer a promising vision for the realization of a hybrid quantum network based on magnonic systems, and find potential applications in quantum-information science and in the study of macroscopic quantum states.

\section*{Acknowledgments}

We thank S. Gr\"oblacher, D. Vitali, and H. Xie for helpful discussions. This work is supported by the National Natural Science Foundation of China (Grants Nos. U1801661, 11934010, 12174329), Zhejiang Province Program for Science and Technology (Grant No. 2020C01019), and the Fundamental Research Funds for the Central Universities (No. 2021FZZX001-02).

\setcounter{figure}{0}
\renewcommand{\thefigure}{A\arabic{figure}}
\setcounter{equation}{0}
\renewcommand{\theequation}{A\arabic{equation}}

\section*{APPENDIX}

In this section, we provide details on the derivation of the solutions in Eq.~\eqref{Aout-m}, which account for the optomagnonic state-swap operation. By integrating the second equation in Eq.~\eqref{a2out-m}, we obtain the following solution
\begin{equation}
m(\tau_2)=e^{-\GG_2 \tau_2} m(0) - i \sqrt{2 \GG_2} \, e^{-\GG_2 \tau_2} \!\! \int_0^{\tau_2} e^{\GG_2 s} a_2^{\rm in} (s)\, ds.
\end{equation}
Using the definition of the temporal mode $A_2^{\rm in} (t)$ provided in Eq.~\eqref{tempo}, the above equation can be rewritten as
\begin{equation}\label{mmmt}
m(\tau_2)=e^{-\GG_2 \tau_2} m(0) - i \sqrt{ 1- e^{-2\GG_2 \tau_2} } A_2^{\rm in}(\tau_2),
\end{equation}
that is the second equation in Eq.~\eqref{Aout-m}.

The second equation of Eq.~\eqref{a2out-m} can also be written in the form of
\begin{equation}
 a_2^{\rm in} = \frac{1}{- i \sqrt{2 \GG_2} } (\dot{m} +\GG_2  m).
\end{equation}
Substituting it into the first equation of Eq.~\eqref{a2out-m}, we obtain 
\begin{equation}
 \dot{m} = \GG_2  m - i \sqrt{2 \GG_2} a_2^{\rm out},
\end{equation}
and its solution is given by 
\begin{equation}
m(\tau_2)=e^{\GG_2 \tau_2} m(0) - i \sqrt{2 \GG_2} \, e^{\GG_2 \tau_2} \!\! \int_0^{\tau_2} e^{-\GG_2 s} a_2^{\rm out} (s)\, ds.
\end{equation}
By using the definition of the temporal mode $A_2^{\rm out} (t)$, we have 
\begin{equation}
m(\tau_2)=e^{\GG_2 \tau_2} m(0) - i \sqrt{ e^{2\GG_2 \tau_2}-1 } A_2^{\rm out}(\tau_2).
\end{equation}
Using the result of Eq.~\eqref{mmmt}, we get the solution of $A_2^{\rm out}(\tau_2)$, which is
\begin{equation}\label{AAAout}
 A_2^{\rm out} (\tau_2) = - i \sqrt{ 1- e^{-2\GG_2 \tau_2} } m(0) + e^{-\GG_2 \tau_2} A_2^{\rm in} (\tau_2).
\end{equation}
Equations~\eqref{mmmt} and \eqref{AAAout} are just the solutions of Eq.~\eqref{Aout-m} in the main text. In the same way, one can derive the solutions of Eq.~\eqref{AAAmmmt1}, which lead to an optomagnonic two-mode squeezed state.


\begin{thebibliography}{99}
\bibitem{note}
Optical pulses with duration of a few tens of ns have been used in the experiments~\cite{Simon16,Simon18}. The time scale is much shorter than the magnon lifetime, typically of 1 $\mu$s. For example, the magnon linewidth $\kappa_m/2\pi= 0.6$ MHz at temperature about 1 K was measured~\cite{S2}, corresponding to lifetime $2\pi/\kappa_m \simeq 1.66$ $\mu$s. Therefore, during the pulse interaction (a few tens of ns) the magnon dissipation is negligible.
\bibitem{notee}
For convenience, we describe the two anti-Stokes processes in optomagnonic and optomechanical systems in independent time reference frames, i.e., $t$ is not unified in the whole process.
\bibitem{nnnot}
For the magnon mode of frequency about 7 GHz~\cite{S7,S8,S9}, the thermal occupation $\bar n \simeq 0.001$ at temperature $T=50$ mK. Therefore, at tens of mK, the magnon mode can be assumed in the ground state.
\end{thebibliography}
\end{document}